%% file: cpv_2012_main_PRD.tex
\documentclass[aps,prd,twocolumn,showpacs,superscriptaddress,groupedaddress,
  ,amsmath,amssymb]{revtex4}  % for review and submission
\usepackage{graphicx}  % needed for figures
\usepackage{dcolumn}   % needed for some tables
\usepackage{bm}        % for math
\usepackage{amssymb}   % for math
\usepackage{multirow} % for multiple rows in tables
\usepackage{graphics}% Include figure files

\input cpv_2012_Symbols.tex

\usepackage[usenames]{color}

\begin{document}

% The following information is for internal review, please remove them for submission
\widetext
%\leftline{Version 1.0 as of \today}
%\leftline{Primary authors: G. Borissov and B. Hoeneisen}
%\leftline{To be submitted to PRD.}
%\leftline{Comment to {\tt d0-run2eb-nnn@fnal.gov} by xxx, yyy}
%\centerline{\em INTERNAL DOCUMENT -- NOT FOR PUBLIC DISTRIBUTION}

% the following line is for submission, including submission to the arXiv!!
\hspace{5.2in} \mbox{Fermilab-Pub-13-055-E}

\title{Understanding the like-sign dimuon charge asymmetry
in $p \bar{p}$ collisions}
\input author_list.tex       % D0 authors (remove the first 3 lines
                             % of this file prior to submission, they
                             % contain a time stamp for the authorlist)
                             % (includes institutions and visitors)
%\author{G. Borissov and B. Hoeneisen}

\date{\today}

\begin{abstract}
The D\O\ collaboration has measured the like-sign dimuon 
charge asymmetry in $p \bar{p}$ collisions at the Fermilab Tevatron
collider. The result is significantly different from the standard model expectation
of CP violation in mixing.
In this paper we consider the possible causes of this asymmetry and
identify one standard model source not considered before.
It decreases the discrepancy of the like-sign dimuon charge asymmetry
with the standard model prediction, although does not eliminate it completely.
\end{abstract}

\maketitle

\input cpv_2012_introduction.tex
\input cpv_2012_composition.tex
\input cpv_2012_source.tex
\input cpv_2012_conclusions.tex

%\section{\label{sec:level1}First-level heading}
% sections are not used for PRL papers

\input cpv_2012_acknowledgement.tex   % input acknowledgement

\end{document}

%% file: cpv_2012_Symbols.tex
\newcommand {\Bd} {\ensuremath{B^0}}
\newcommand {\Bs} {\ensuremath{B^0_s}}
\newcommand {\Bq} {\ensuremath{B^0_q}}

\newcommand {\barBd} {\ensuremath{\bar{B}^0}}
\newcommand {\barBs} {\ensuremath{\bar{B}^0_s}}
\newcommand {\barBq} {\ensuremath{\bar{B}^0_q}}
\newcommand {\asld} {\ensuremath{a^d_{\mathrm{sl}}}}
\newcommand {\asls} {\ensuremath{a^s_{\mathrm{sl}}}}
\newcommand {\aslq} {\ensuremath{a^q_{\mathrm{sl}}}}

%% file: author_list.tex
\affiliation{Lancaster University, Lancaster LA1 4YB, United Kingdom}
\affiliation{Universidad San Francisco de Quito, Quito, Ecuador}
\author{G.~Borissov} \affiliation{Lancaster University, Lancaster LA1 4YB, United Kingdom}
\author{B.~Hoeneisen} \affiliation{Universidad San Francisco de Quito, Quito, Ecuador}

%% file: cpv_2012_introduction.tex
\section{Introduction}
\label{Introduction}

The D\O\ Collaboration has measured \cite{D01,D02,D02a,D03} 
the like-sign dimuon charge asymmetry
and the inclusive muon charge asymmetry
in $p\bar{p}$ collisions at a center-of-mass
energy $\sqrt{s}=1.96$~TeV at the Fermilab Tevatron collider.
After subtracting the known background sources, the inclusive muon 
charge asymmetry is found to be compatible with zero
while the like-sign dimuon charge asymmetry significantly deviates from 
the standard model (SM) prediction.
%In Refs. \cite{D01,D02,D02a,D03} 
This deviation is usually interpreted as the
anomalous charge asymmetry of $\Bd$ and $\Bs$ semileptonic decays \cite{Grossman, lenz}.

In this paper we consider other possible sources of the like-sign dimuon charge asymmetry,
taking into account the constraint that the inclusive muon asymmetry is consistent with zero.
We identify one significant contribution which
was not accounted for previously. In addition, we discuss all other sources
of the dimuon charge asymmetry and show that the measurements of the D\O\ collaboration
put strong constraints on them. After presenting available experimental results
related to the dimuon charge asymmetry in Section II, we consider in Section III one by one 
the contribution of different processes into this asymmetry. Our results are summarised in 
Section IV. 

%From the information presented in \cite{D03} we obtain
%\begin{eqnarray}
%A_S & = & (-0.383 \pm 0.092\textrm{(stat)} \pm 0.102\textrm{(syst)})\%, \label{Ac3} \\
%a_S & = & (-0.063 \pm 0.079\textrm{(stat)} \pm 0.141\textrm{(syst)})\%. \label{ac3}
%\end{eqnarray}
%((\ref{Ac3}) is obtained by dividing (35) of \cite{D03} by
%$C_b = 0.474 \pm 0.032$ \cite{D03}.)
%The striking observation is that $a_S$ is compatible with zero,
%while $A_S$ is significantly different from zero.
%The purpose of this publication is to find the source of the measured
%like-sign dimuon charge asymmetry (\ref{Ac3}) subject to the
%constraint (\ref{ac3}).
%

%% file: cpv_2012_composition.tex
\section{Experimental results}
\label{composition}

We use the results of Ref. \cite{D03} and express them in a model independent way
as the charge asymmetry of the inclusive muon sample $a_S$, and the charge asymmetry of
the like-sign dimuon sample $A_S$. We follow the notations and definitions of Ref. \cite{D03},
where the asymmetries $a_S$ and $A_S$ are computed using the muons coming from the
decays of $b$ and $c$ quarks and $\tau$ leptons, and from decays of short-living
mesons such as $\phi$, $\omega$, $\eta$, $\rho^0$, $J/\psi$. The origin 
of the asymmetries $a_S$ and $A_S$ may be the semileptonic charge asymmetry of $\Bd$ and $\Bs$
decays, as well as other $CP$-violating processes.

From the information presented in \cite{D03} we obtain
\begin{eqnarray}
a_S & = & (-0.063 \pm 0.079\textrm{(stat)} \pm 0.141\textrm{(syst)})\%, \label{ac3} \\
A_S & = & (-0.383 \pm 0.092\textrm{(stat)} \pm 0.102\textrm{(syst)})\%. \label{Ac3}
\end{eqnarray}
These values are obtained, respectively, by multiplying the results given in Eqs. (34) and (35) 
by the coefficients $c_b$ and $C_b$ given in Eqs. (30) and (31) of Ref. \cite{D03}. 
An important observation which can be derived from these results 
is that $a_S$ is compatible with zero,
while $A_S$ is significantly different from zero.

The composition of the inclusive muon sample is presented, for convenience, in
Table \ref{comp} taken from Ref. \cite{D03}. 
The composition of the like-sign dimuon sample can be derived from the information given in
Table \ref{comp} assuming that the two muons come from independent processes.
This assumption can be applied because the requirement that the
invariant mass of the two like-sign muons is greater than 2.8 GeV, which is used for selecting
the dimuon pairs \cite{D03}, suppresses decays with 
the two muons originating from the same $b$ quark.
%The contribution of the process $T_6$ in Table \ref{comp} is excluded from 
%the like-sign dimuon sample. 
Since the oscillation of the $D^0$ meson is found to be small \cite{PDG}, 
the contribution of process
$T_6$ in Table \ref{comp} to the like-sign dimuon sample is suppressed and,
therefore, is neglected in the following discussion. 

%In the inclusive muon sample, but excluding processes $T_6$,
%the probability that an
%initial $b$ of $\bar{b}$ quark decay to a ``right sign"
%muon (i.e. to a muon with charge of the same sign as the
%charge of the initial $b$ of $\bar{b}$ quark at the time
%of the $p \bar{p}$ collision) is proportional to
%$P_b = w_{1a} + w_{2b} +0.5 (w_3 + w_4 + w_5) = 0.934 \pm 0.005$,
%while the probability to decay to a ``wrong sign" muon
%is proportional to
%$\bar{P_b} = w_{1b} + w_{2a} +0.5 (w_3 + w_4 + w_5) = 0.260 \pm 0.012$. \cite{D02, D03}

\begin{table}
\caption{\label{comp}
Heavy-quark decays contributing to the inclusive muon and like-sign dimuon
samples taken from Ref. \cite{D03}.
The abbreviation ``non-osc" stands for ``non-oscillating," and ``osc" for ``oscillating."
All weights are computed using MC simulation. $\chi_0 = 0.1260 \pm 0.0037$ is the time-integrated 
mixing probability \cite{hfag}.
}
\begin{ruledtabular}
\newcolumntype{A}{D{A}{\pm}{-1}}
\newcolumntype{B}{D{B}{-}{-1}}
\begin{tabular}{lll}
   & Process & Weight \\
\hline
$T_1$   & $b \to \mu^-X$ & $w_1 \equiv 1.$ \\
$T_{1a}$ & ~~~$b \to \mu^-X$ (non-osc) & $w_{1a} = (1-\chi_0) w_1$ \\
$T_{1b}$ & ~~~$\bar{b} \to b \to \mu^-X$ (osc) & $w_{1b} = \chi_0 w_1$ \\
$T_2$ &  $b \to c \to \mu^+X$ & $w_2 = 0.096 \pm 0.012$ \\
$T_{2a}$ & ~~~$b \to c \to \mu^+X$ (non-osc) & $w_{2a} = (1-\chi_0) w_2$ \\
$T_{2b}$ & ~~~$\bar{b} \to b \to c \to \mu^+X$ (osc) & $w_{2b} = \chi_0 w_2$ \\
$T_3$ & $b \to c \bar c q$ with $c \to \mu^+X$ or $\bar c \to \mu^-X$ & $w_3 = 0.064 \pm 0.006$ \\
$T_4$ & $\eta, \omega, \rho^0, \phi(1020), J/\psi, \psi' \to \mu^+ \mu^-$ & $w_4 = 0.021 \pm 0.002$ \\
$T_5$ & $b \bar b c \bar c$ with $c \to \mu^+X$ or $\bar c \to \mu^-X$ & $w_5 = 0.013 \pm 0.002$ \\
$T_6$ & $c \bar c$ with $c \to \mu^+X$ or $\bar c \to \mu^-X$ & $w_6 = 0.675 \pm 0.101$
\end{tabular}
\end{ruledtabular}
\end{table}

For our purposes it is convenient to
classify the processes $T_3$ into three different categories denoted $T_{3d}(f_i)$,
$T_{3s}(f_j)$ and $T_{3 {\textrm{fs}}}$.
Processes $T_{3d}(f_i)$ correspond to the decays of 
$\Bd$ or $\barBd$ to the final state $f_i$ containing two $c$ quarks
and accessible to both $B^0$ and
$\bar{B}^0$. At the quark level these decays correspond to the process
$(\bar b d)$ or $(b \bar d) \to c \bar c d \bar d$. 
Similarly, processes $T_{3s}(f_j)$	correspond to the decays of $\Bs$
or $\barBs$ to the final state $f_j$ containing two $c$ quarks and
accessible to both $\Bs$ and $\barBs$. These decays at the quark level
are $(\bar b s)$ or $(b \bar s) \to c \bar c s \bar s$.
All other decays of $B$ hadrons producing two charm quarks
are flavour specific and included in the group denoted $T_{3 {\textrm{fs}}}$.
The weights of these processes are respectively
$w_{3d}(f_i)$, $w_{3s}(f_j)$ and $w_{3 {\textrm{fs}}}$. By definition,
\begin{equation}
w_3 = \sum_i w_{3d}(f_i) + \sum_j w_{3s}(f_j) + w_{3 {\textrm{fs}}}.
\end{equation}

%% file: cpv_2012_source.tex
\section{Contributions to the like-sign dimuon charge asymmetry}
\label{source}

We consider in this section some SM processes producing
the like-sign dimuon charge asymmetry $A_S$, and the current experimental
constraints of their contribution taken from \cite{PDG}.
%(we also adopt the notation of \cite{PDG} that is in general use).
Particles of physics beyond the SM may add new Feynman diagrams
with loops in $B^0 \leftrightarrow \bar{B}^0$ mixing,
$\Bs \leftrightarrow \barBs$ mixing, and
in penguin decays. They are not discussed here.

The main source of the like-sign dimuon pairs produced
in $p \bar p$ collisions is $b \bar b$ events.
The background muons from $K \to \mu \nu$ and $\pi \to \mu \nu$ decays
are excluded by definition from the asymmetries $a_S$ and $A_S$, and are not
discussed here. One of the $B$ hadrons from the $b \bar b$ pair
%Consider the following sequence of events:
%the $p \bar{p}$ collision produces a $b \bar{b}$ pair
%that hadronizes into two $b$-hadrons. One of these
%$b$-hadrons
decays to a ``right sign" muon, i.e.
to a muon with charge of the same sign as
the charge of the initial $b$ or $\bar{b}$ quark
at the time of the $p \bar{p}$ collision. To obtain
a like-sign dimuon event, the other $b$-hadron must
decay to a ``wrong sign" muon. 
%In the following
%we mainly consider $CP$ violation (CPV)
%of decays to a ``wrong sign" muon.
%The $CP$ violation of decays to a ``right sign"
%muon is constrained by the ``closure test" (\ref{ac3})
%as explained in Subsection \ref{semileptonic} below.

\subsection{$CP$ violation in mixing of $\Bd$ and $\Bs$ mesons}
The complex phase $\phi_{12}^q$ of the mass matrix of the $\Bq$ $(q=d,s)$ system
produces the charge asymmetry $\aslq$ of the ``wrong sign" semileptonic $\Bq$ decays defined as
\begin{equation}
\aslq \equiv \frac{\Gamma(\barBq(t)\rightarrow l^+ X) -
              \Gamma(   \Bq(t)\rightarrow l^- X)}
             {\Gamma(\barBq(t)\rightarrow l^+ X) +
              \Gamma(   \Bq(t)\rightarrow l^- X)}.
              \label{aslq}
\end{equation}
The asymmetry $\aslq$ is related to the phase $\phi_{12}^q$ as \cite{PDG}
\begin{equation}
\aslq = \frac{\Delta \Gamma_q}{\Delta m_q} \tan \phi^q_{12}.
\label{aslq_tan}
\end{equation}
The contributions of this asymmetry to the inclusive muon charge asymmetry
and the like-sign dimuon charge asymmetry are
\begin{eqnarray}
a_S(\mbox{from \aslq}) & = & c_b C_q \aslq, \\
A_S(\mbox{from \aslq}) & = & C_b C_q \aslq,
\end{eqnarray}
where the coefficients $c_b$, $C_b$, $C_d$ and $C_s$ are defined in Ref. \cite{D03}:
\begin{eqnarray}
c_b = 0.061 \pm 0.007, & ~~~ & C_b  =  0.474 \pm 0.032, \\
C_d = 0.594 \pm 0.022, & ~~~ & C_s  =  0.406 \pm 0.022.
\end{eqnarray}

The SM predictions of the phases $\phi^q_{12}$ and asymmetries $\aslq$ are \cite{lenz}:
\begin{eqnarray}
\phi^d_{12}\mbox{(SM)} & = & -0.075 \pm 0.024, \\
\phi^s_{12}\mbox{(SM)} & = & +0.0038 \pm 0.0010,  \\
\asld \mbox{(SM)} & = & -(4.1 \pm 0.6) \times 10^{-4}, \\
\asls \mbox{(SM)} & = & +(1.9 \pm 0.3) \times 10^{-5}.
\end{eqnarray}

%The SM prediction can not explain the observed value of the $A_S$. The contribution
%of new physics can significantly modify the phase $\phi_{12}^s$ resulting in a larger
%value of $A_S$.
Recently, the experimental measurements of both $\asld$ and $\asls$ became available.
The measurements of $\asld$ are performed at $\Upsilon(4S)$ \cite{hfag} 
and by the D\O\ experiment \cite{asld-d0}:
\begin{eqnarray}
\asld & = & +0.0002 \pm 0.0031 ~~(\Upsilon\mbox{(4S)}), \\
\asld & = & +0.0068 \pm 0.0047 ~~\mbox{(D\O)}.
\end{eqnarray}
Our combination of these values gives
\begin{equation}
\asld = +0.0022 \pm 0.0026.
\label{asld1}
\end{equation}

The measurements of $\asls$ are performed by D\O\ \cite{asls-d0} and LHCb \cite{asls-lhcb} collaborations:
\begin{eqnarray}
\asls & = & -0.0104 \pm 0.0074 ~~\mbox{(D\O)}, \\
\asls & = & -0.0024 \pm 0.0063 ~~\mbox{(LHCb)}.
\end{eqnarray}
Our combination of these values gives
\begin{equation}
\asls = -0.0058 \pm 0.0048.
\label{asls1}
\end{equation}

Using the values (\ref{asld1}) and (\ref{asls1}) we obtain the
allowed contributions to the
inclusive muon and
like-sign dimuon charge asymmetries from CPV in mixing of $\Bd$ and $\Bs$ mesons:
\begin{eqnarray}
a_S(\mbox{from \asld}) & = & (+0.008 \pm 0.009) \%, \\
A_S(\mbox{from \asld}) & = & (+0.062 \pm 0.073) \%, \\
a_S(\mbox{from \asls}) & = & (-0.014 \pm 0.012) \%, \\
A_S(\mbox{from \asls}) & = & (-0.111 \pm 0.093) \%, 
\end{eqnarray}
and their sum
\begin{eqnarray}
a_S(\mbox{from \aslq}) & = & (-0.006 \pm 0.015) \%, \\
A_S(\mbox{from \aslq}) & = & (-0.049 \pm 0.118) \%.
\end{eqnarray}
For comparison, the SM prediction is
\begin{equation}
A_S(\mbox{from {\aslq} in SM})  =  (-0.013 \pm 0.002) \%.
\label{SM_B0_mix}
\end{equation}

In addition, the estimate of $\asls$ can be extracted from the measurement
of $CP$ violation in the $\Bs \to J/\psi \phi$ decay. The corresponding phase
$\phi_s^{c \bar c s}$ is expected to change in the same way as the phase $\phi_{12}^s$
due to a new physics contribution $\phi_s^{\Delta}$ \cite{Nierste,lenz}:
\begin{eqnarray}
\phi_s^{c \bar c s} & = & \phi_s^{c \bar c s} \mbox{(SM)} + \phi_s^{\Delta},  \label{delta} \\
\phi_{12}^s & = & \phi_{12}^s \mbox{(SM)} + \phi_s^{\Delta}, \nonumber \\
\phi_s^{c \bar c s} \mbox{(SM)} & = & - \sin (2 \beta_s) = -0.036 \pm 0.002.
\end{eqnarray}
The current world average value of $\phi_s^{c \bar c s}$ is \cite{hfag}:
\begin{equation}
\phi_s^{c \bar c s} = -0.013^{+0.083}_{-0.090}.
\end{equation}
From these expressions we get the following estimate of $\asls$ from $\Bs \to J/\psi \phi$ decay:
\begin{equation}
\asls (\mbox{from}~ \Bs \to J/\psi \phi)= -0.0001 \pm 0.0005.
\label{asls2}
\end{equation}
However, we do not use the result (\ref{asls2})
because Eq. (\ref{delta}) 
may be subject to large penguin corrections from new physics \cite{lenz2}.

%The decay $B_s \rightarrow J/\psi \phi$ at LHCb obtains
%$\phi_s + \delta^\textrm{peng.}_s = -0.001 \pm 0.101 \textrm{ (stat)} \pm 0.027 \textrm{ (syst)}$
%\cite{LHCb} {\color{red}(do we have a better reference?)},
%where $\delta^\textrm{peng.}_s$ is a penguin diagram
%contribution from the SM and new physics. Assuming
%$\delta^\textrm{peng.}_s$ is negligible, we obtain
%$a^s_\textrm{sl} = (0.00 \pm 0.07)\%$. The contribution
%of $a^s_\textrm{sl}$ to $A_S$ is
%$C_b C_s a^s_\textrm{sl} = (0.000 \pm 0.014)\%$,
%where $C_b = 0.474 \pm 0.032$ and $C_s = 0.406 \pm 0.022$. \cite{D03}
%In conclusion,
%unless there is a cancellation of $\phi_s$ and $\delta^\textrm{peng.}_s$,
Thus,
%assuming the validity of the assumption
%(\ref{delta}), $CP$ violation in mixing of $\Bd$ and $\Bs$ mesons cannot explain the
%measured like-sign dimuon charge asymmetry.
%
%\subsection{CPV in mixing of $B^0$.}
%\textbf{2. CPV in mixing of $B^0$.}
%Assuming $a^s_\textrm{sl}$ is negligible, we obtain
%from Figure 21 of \cite{D03}
%\begin{equation}
%a^d_\textrm{sl} = (-0.78 \pm 0.23 \textrm{ (tot)})\%
%\label{adsl3}
%\end{equation}
%at 68\% CL. {\color{Red}(this number needs to be updated)}
%
%The world average semileptonic asymmetry from this source
%is $a^d_\textrm{sl} = (-0.10 \pm 0.30)\%$. \cite{WA}
%The contribution of $a^d_\textrm{sl}$ to $A_S$ is
%$C_b C_d a^d_\textrm{sl} = -0.028 \pm 0.094\%$, where
%$C_d = 0.594 \pm 0.022$ \cite{D03},
%so
the current experimental results
constrain the contribution of
CPV in mixing of $\Bd$ and $\Bs$ mesons to the 
measured like-sign dimuon charge asymmetry.
%Global fits by the \textsf{CKMfitter} group \cite{CKMfitter}, with new physics
%only in the matrix elements $M^d_{12}$ and $\Gamma^d_{12}$, are
%still viable. Note that CPV is required in the
%matrix element $\Gamma^d_{12}$, which is unexpected in the simplest
%extensions of the SM.

\subsection{$CP$ violation in interference of $\Bd$ decay with and
without mixing}
% to $f_i$, followed by $f_i \rightarrow \mu^- X$.}
%\textbf{3. CPV in interference of decay with and
%without mixing of $B^0$, followed by $\bar{c} \rightarrow \mu^- X$.}
The final states of the decay $\Bd (\barBd) \to c \bar c d \bar d$
are accessible from both $\Bd$ and $\barBd$. Therefore, the interference of decays
to these final states with and without $\Bd$ mixing results in CPV \cite{PDG}.
It turns out that this CPV produces a like-sign dimuon charge asymmetry. At the
same time, its contribution is negligible in the inclusive muon charge asymmetry.

To demonstrate this, let us consider an example of the process
producing a positive dimuon pair
\begin{eqnarray}
p \bar{p} & \rightarrow & B^+ \barBd X,  \nonumber \\
B^+ & \rightarrow & \mu^+ X, \nonumber \\
\barBd  & \rightarrow & D^+ D^-, D^+ \rightarrow \mu^+ X.
\end{eqnarray}
%The $B^+$ meson decays semileptonically $B^+ \to \mu^+ X$.
The state $f_i = D^+ D^-$ is a CP-even
eigenstate accessible from both $B^0$ and $\bar{B}^0$ mesons.
The $D^+ \rightarrow \mu^+ X$ decay produces
a ``wrong sign" muon, and contributes to the
like-sign dimuon sample.
The $D^- \rightarrow \mu^- X$ decay produces
a ``right sign" muon, and therefore does not contribute
to the like-sign dimuon sample.
The number of positive and negative muons from the decay
$\Bd \to D^+ D^-$ is the same, therefore there is no contribution
to the inclusive muon charge asymmetry from this decay.

%Note that in the inclusive
%muon sample, the muon charges of $D^-	\rightarrow \mu^- X$ and
%$D^+	\rightarrow \mu^+ X$ cancel, so \textit{CPV in interference only
%produces a charge asymmetry in the like-sign dimuon sample, not
%in the inclusive muon sample}.
The decay rate of the meson that is initially produced as a $\barBd$ is \cite{PDG}
\begin{eqnarray}
& & \frac{d \Gamma(\barBd \to f_i)}{dt}  \propto  \exp{(-\Gamma_d t)} \times \nonumber \\
\label{dGdt}
& & \left[ 1 + S_i \sin{(\Delta m_d t)} - C_i \cos{(\Delta m_d t)} \right].
\end{eqnarray}
The term proportional to $S_i$ is due to CPV in the interference
of decays with and without mixing to $D^+ D^-$,
of the meson that is initially produced as $\bar{B}^0$.
The term proportional to $C_i$ is due to the dirct CPV in $\Bd$ decay.
Neglecting the loop contributions to the decay amplitude, 
the coefficients $S_i$ and $C_i$ in the SM are expressed as \cite{Sanda}
\begin{eqnarray}
S_i & = & -\eta_i \sin (2 \beta), \\
C_i & = & 0, \nonumber
\end{eqnarray}
where $\eta_i$ is the $CP$ eigenvalue of the final state $f_i$, and $\beta$ is the
angle of the Unitarity Triangle \cite{PDG}. 
The loop diagrams can change this estimate by a few percent \cite{Xing1,Xing2}.
For the $\Bd \to D^+ D^-$ final state $\eta_i = +1$.
%If $f_i$ is an eigenstate of CP, then
%$S_i = \eta_i \sin{(2 \beta)} = 0.679 \pm 0.020$, and
%$\eta_i = +1$ or $-1$ is the CP eigenvalue.
If $f_i$ is not a $CP$ eigenstate, the coefficient $C_i$ may be non-zero,
but experimentally for the decays considered here
$C_i$ is negligible \cite{PDG}. Therefore, we omit it in the following discussion.
Integrating (\ref{dGdt}) we obtain the width of the decay to this final state
\begin{eqnarray}
\Gamma(\barBd \to f_i) & \propto & 1 + \frac{S_i x_d}{1 + x_d^2}, \label{G} \\
x_d & \equiv & \frac{\Delta m_d} {\Gamma_d}. \nonumber
\end{eqnarray}
Now consider the CP-conjugate process producing a negative dimuon pair
\begin{eqnarray}
p \bar{p} & \rightarrow & B^- \Bd X, \nonumber \\
 B^- & \rightarrow & \mu^- X, \nonumber \\
\Bd  & \rightarrow & D^+ D^-, D^- \rightarrow \mu^- X.
\end{eqnarray}
The decay rate of the meson that is initially produced as a $\Bd$ is
\begin{equation}
\frac{d \Gamma (\Bd \to f_i)}{dt} \propto \exp{(-\Gamma_d t)} \left[ 1 - S_i \sin{(\Delta m_d t)} \right],
\label{dGdt_bar}
\end{equation}
and the partial width is
\begin{equation}
\Gamma (\Bd \to f_i) \propto 1 - \frac{S_i x_d}{1 + x_d^2}.
\label{G_bar}
\end{equation}
The like-sign dimuon charge asymmetry from this process is
\begin{equation}
A_i = S_i \frac{x_d}{1 + x_d^2}.
\label{asym}
\end{equation}
Numerically the absolute value of this asymmetry is large, because $\sin (2 \beta) = 0.679 \pm 0.020$
and $x_d = 0.770 \pm 0.008$ \cite{PDG}. 

Let us now obtain the contribution of the decay channel
$f_i$ with weight $w_{3d}(f_i)$ to the like-sign
dimuon charge asymmetry $A_S$. 
This weight takes into account both the branching fraction
of $\Bd \to f_i \to \mu X$ decay and detector-related efficiency of 
muon reconstruction. The weights for the
various processes have been defined in Section \ref{composition}.
The probability that an initial $b$ quark produces
a ``right sign" muon $\mu^-$ is \cite{D01}
\begin{eqnarray}
P_b & \propto & 0.5 w_{3d}(f_i)
  \left[ 1 + S_i \frac{x_d}{1 + x_d^2} \right]
  + w_{1a} \nonumber \\
& & + w_{2b} + 0.5 (w_{3s} + w_{3\textrm{fs}} + w_4 + w_5).
\label{Pb}
\end{eqnarray}
The factor 0.5 in this expression corresponds to the statement that the
number of positive and negative muons produced in the processes $T_3$, $T_4$ and $T_5$
is the same. The probability that an initial $\bar{b}$ quark produces a
``wrong sign" muon $\mu^-$ is
\begin{eqnarray}
P_{\bar{b}} & \propto & 0.5 w_{3d}(f_i)
  \left[ 1 - S_i \frac{x_d}{1 + x_d^2} \right]
  + w_{1b} \nonumber \\
& & + w_{2a} + 0.5 (w_{3s} + w_{3\textrm{fs}} + w_4 + w_5).
\label{Pb_bar}
\end{eqnarray}
The number of $\mu^- \mu^-$ events is proportional to
$P_b P_{\bar{b}}$.
The number of $\mu^+ \mu^+$ events is obtained by
replacing $S_i$ by $-S_i$. The charge asymmetry
from channels $f_i$ is then
\begin{equation}
A_S(f_i) = 0.5 w_{3d}(f_i) S_i \frac{x_d}{1 + x_d^2}
\frac{P_b - P_{\bar{b}}}{P_b P_{\bar{b}}}.
\label{Ad_fi}
\end{equation}

Thus, CPV in interference of $\Bd$ decay with and without mixing
produces a like-sign dimuon charge asymmetry, while it does not contribute
to the inclusive muon charge asymmetry. The possible final states 
produced in the $\Bd (\barBd) \to c \bar c d \bar d$ decay include
$D^+ D^-$, $D^{*+} D^-$, $D^+ D^{*-}$, $D^{*+} D^{*-}$, $J/\psi \pi^0$,
$J/\psi \eta$, $J/\psi \rho^0$, etc. For many of these final states the value $S_i$
is measured experimentally \cite{PDG} and in all cases it is consistent with the
SM value $S_i = - \sin (2 \beta)$, which corresponds to the expectation of the dominance
of the $CP$-even final states in the $\Bd (\barBd) \to c \bar c d \bar d$ decay.

On the contrary, the contribution of CPV in the $b \to c \bar c s$ transition
producing the $CP$ eigenstates, like the $\Bd \to J/\psi K_S$ decay, should not contribute
to the like-sign dimuon charge asymmetry, because for each $CP$-even final
state there should be the corresponding $CP$-odd final state. For example, the contribution from the decay
$\Bd \to J/\psi K_S$ is canceled by the decay $\Bd \to J/\psi K_L$.

The weight $w_{3d}(f_i)$ can be obtained by 
%counting simulated
%events, or alternatively, by 
using the measured branching fraction of
$\Bd \to f_i$ decay. For example, the weight $w_{3d}(D^+D^-)$
can be computed using the following expression
\begin{equation}
w_{3d}(f_i) = f_d \alpha \frac{\textrm{Br}(\Bd \rightarrow D^+ D^-)
  2 \textrm{Br}(D^+ \rightarrow \mu X)}{\textrm{Br}(b \rightarrow \mu X)}.
\label{w3d}
\end{equation}
The coefficient $w_{3d}(f_i)$ by definition is normalised to the weight $w_1$,
which is proportional to $\textrm{Br}(b \rightarrow \mu X)$.
The value $\textrm{Br}(b \rightarrow \mu X)$ is the average branching fraction
of the direct decay of all $B$ hadrons to muon weighted with the
relative production rate of different $B$ hadrons at the hadron collider. 
Also by definition, the weight $w_{3d}(f_i)$ includes both
decays $D^+ \rightarrow \mu^+ X$ and $D^- \rightarrow \mu^- X$,
hence the factors 0.5 in Eqs. (\ref{Pb}) to (\ref{Ad_fi}), and the factor 2 in
Eq. (\ref{w3d}). The factor $f_d = 0.401 \pm 0.008$ \cite{PDG} is
the fraction of $\Bd$ plus $\barBd$ in the admixture of $b$-hadrons.
%Following the definitions of Ref. \cite{D03}, this branching fraction also includes
%the contribution from $b \to \tau X \to \mu X$ decay.
%The factor $\alpha = 0.243 \pm 0.047 \textrm{(stat)} \pm 0.020 \textrm{(syst)}$
%is the ratio of detector acceptances
%of muons from $D^+$ and $\Bd$ decay, and is obtained form the simulation.
The factor $\alpha$ is the ratio of detector acceptances
of muons from $D^+$ and $\Bd$ decay.
Muons from $D^+$ and $\Bd$ mesons have different detector acceptances
because they have different kinematic distributions.
%From the simulation of $\Bd \rightarrow \mu^\pm X$
%and $\Bd \rightarrow D^+ D^- \rightarrow \mu^\pm X$ we obtain
%$\alpha = 0.244 \pm 0.057 \textrm{(stat)}$; from the simulation of
%$\Bs \rightarrow \mu^\pm X$
%and $\Bs \rightarrow D^+_s D^-_s \rightarrow \mu^\pm X$ we obtain
%$\alpha	= 0.242	\pm 0.082 \textrm{(stat)}$.

Using the results of Ref. \cite{D03} and other experimental values from \cite{PDG}
we estimate the coefficient $\alpha$ from the following expressions:
\begin{eqnarray}
w_1 & \propto & \textrm{Br}(b \rightarrow \mu X), \nonumber \\
w_3 & \propto & \textrm{Br}(b \rightarrow c \bar c X) 
\textrm{Br}(c \bar c q \bar q' \rightarrow \mu X) \alpha.
\end{eqnarray}
Here $\textrm{Br}(b \rightarrow c \bar c X)$ is the branching fraction
of $B$ hadron decays producing $c \bar c$ pair. We use the experimental value
$\textrm{Br}(b \rightarrow c \bar c X) = 0.162 \pm 0.032$ which is obtained
from $\textrm{Br}(B^\pm / \Bd / \Bs / b\mbox{-baryon mixture} \to \bar c / c X) = 1.162 \pm 0.032$  \cite{PDG}.
The quantity $\textrm{Br}(c \bar c q \bar q' \rightarrow \mu X)$ is the average branching fraction
of the direct decay of all charmed hadrons to muon weighted with the
relative production rate of different pairs of $c$ hadrons in $B$ hadron decay.
Using the values of corresponding branching fractions for the
$B^\pm / \Bd / \Bs / b$-baryon mixture from Ref. \cite{PDG} we
obtain
\begin{eqnarray}
\textrm{Br}(b \rightarrow \mu X) = 0.107 \pm 0.003, \\
\textrm{Br}(c \bar c q \bar q' \rightarrow \mu X) = 0.164 \pm 0.032.
\end{eqnarray}
From these expressions we obtain
\begin{eqnarray}
\label{alpha}
\alpha & = & w_3 \frac{\textrm{Br}(b \rightarrow \mu X)}
{ \textrm{Br}(b \rightarrow c \bar c X) 
\textrm{Br}(c \bar c q \bar q' \rightarrow \mu X)} \\
& = & 0.258 \pm 0.073. \nonumber
\end{eqnarray}
This estimate of $\alpha$ does not take into account different
kinematic distributions for the various decays $c \rightarrow \mu X$.
Therefore, a simulation of the D\O\ detector is required to obtain
a more accurate value for $\alpha$.

Using Eqs. (\ref{w3d}) and (\ref{alpha}) we obtain
\begin{equation}
w_{3d}(f_i) = f_d w_3 
\frac{\textrm{Br}(\Bd \rightarrow f_i)} {\textrm{Br}(b \to c \bar c X)}
\frac{\textrm{Br}(f_i \rightarrow \mu X)} {\textrm{Br}(c \bar c q \bar q' \rightarrow \mu X)}.
\label{w3d-1}
\end{equation}

Using this observation, we estimate $A_S$ by several methods.

{\bf Estimate 1.}
Let us consider four measured decay channels \cite{PDG}:
$D^+ D^-$ with $S = -0.87 \pm 0.26$, $\Gamma_i/\Gamma = (2.11 \pm 0.31) \times 10^{-4}$;
$D^*(2010)^+ D^-$ with $S = -0.61 \pm 0.19$, $\Gamma_i/\Gamma = (6.1 \pm 1.5) \times 10^{-4}$;
$D^*(2010)^- D^+$ with $S = -0.78 \pm 0.21$, $\Gamma_i/\Gamma \approx 6.1 \times 10^{-4}$ (our guess); and
$D^{*+} D^{*-}$ with $S = -0.76 \pm 0.14$, $\Gamma_i/\Gamma = (8.2 \pm 0.9) \times 10^{-4}$.
Using these numbers we obtain for the sum of these channels
\begin{eqnarray}
& & \sum_{i} (S_i \textrm{Br}(\Bd \rightarrow f_i) \textrm{Br}(f_i \rightarrow \mu X)) = \nonumber \\
         & = & -0.00044 \pm 0.00011.
\end{eqnarray}
Using this value and Eqs. (\ref{Ad_fi}) and (\ref{w3d-1}), 
the contribution to $A_S$ from these 4 channels is
\begin{equation}
%A_S(\textrm{4 channels of }\Bd) = (-0.037 \pm 0.012)\%.
A_S(\textrm{4 channels}) = (-0.028 \pm 0.011)\%.
\label{ASd}
\end{equation}

%{\color{Red}Other channels that should be considered:
%$J/\psi (nS) K^0$ with $S = +0.676 \pm 0.021$, $\Gamma_i/\Gamma = (8.74 \pm 0.32) \times 10^{-4}$; and
%$J/\psi (1S) \pi^0$ with $S = -0.94 \pm 0.29$, $\Gamma_i/\Gamma = (1.76 \pm 0.16) \times 10^{-5}$.
%Other channels are $b \rightarrow \bar{c} c s$, $b \rightarrow \bar{s} s d$, etc.}

{\bf Estimate 2.}
%
%Si = eta * sin(2beta)
%
%Br( Bd -> c cbar d dbar) / Br( b -> c cbar q) = fd * (Vcd)^2
%
%As = 0.5 *w3 * fd * (Vcd)^2 * sin(2 beta) * x/(1+x^2) * (Pb + Pbbar)
%/ (Pb * Pbbar)
%
%x = Delta m / Gamma = 0.77
%sin(2 beta) = 0.67
%Vcd = 0.23
%fd    = 0.40
%
%If I use w3 = 0.064, and other values from Table 1, I get:
%
%As = 0.00109
%
%If $f_i$ is a CP eigenstate with eigenvalue $\eta_i = \pm 1$,
%and if only a single weak phase dominates the decay, and if
%$|\Gamma_{12} / M_{12}| = 0$ (a good approximation for both
%$B^0$ and $B_s$), then $S_i = \eta_i \sin{(2 \beta)}$.
For this estimate we assume that the final states $\bar{c} d c \bar{d}$
are mostly $CP$-even ($\eta_i = +1$), which is appropriate for 
$D^{(*)+} D^{(*)-}$ final states
and confirmed by the experimental measurements.
Then 
\begin{equation}
\frac{\textrm{Br}(\Bd \rightarrow c \bar{c} d \bar{d})}
{(\textrm{Br}(b \rightarrow c \bar{c} s) +
 \textrm{Br}(b \rightarrow c \bar{c} d))}
\approx V^2_{cd}.
\end{equation}
Eq. (\ref{Ad_fi}) becomes approximately
\begin{eqnarray}
A_S & \approx & -0.5 w_3 f_d V_{cd}^2 \sin{(2 \beta)}
\frac{x_d}{1 + x_d^2}
%\frac{\Delta m_d \Gamma_d}{\Gamma_d^2 + \Delta m_d^2}
\frac{P_b - P_{\bar{b}}}{P_b P_{\bar{b}}} \delta, \nonumber \\
\delta & \equiv & \frac{\mbox{Br}(c \bar c d \bar d \to \mu X)}{\mbox{Br}(c \bar c q \bar q' \to \mu X)}.
\end{eqnarray}
The factor $\delta$ takes into account the fact that the final state
$c \bar c d \bar d$ contains more $D^\pm$ mesons than the generic 
$c \bar c q \bar q'$ state, and that the semileptonic branching fraction 
of $D^\pm$ meson is about 2.7 times larger than that of all other charm hadrons.
Using the known branching fractions of $B$-meson decays, we estimate 
$\delta = 1.5 \pm 0.2$. Using the values from Table \ref{comp} we obtain 
the following estimate of $A_S$ from CPV in interference:
\begin{equation}
A_S =  (-0.089 \pm 0.015)\%.
\label{Ad_fi_2}
\end{equation}
This value gives an upper bound of the $-A_S$ estimate, since in deriving it 
we assume that all $c \bar c d \bar d$ states are $CP$-even, which is clearly not the case.

{\bf Estimate 3.}
In the SM the $CP$ violation in mixing of neutral $B$ mesons is small
and the mass eigenstates of the $\Bd$ system coincide with $CP$ eigenstates. 
In that case \cite{Dunietz}
\begin{equation}
\Delta \Gamma = \Delta \Gamma_{CP} = \Gamma (B_d^{0, even}) - \Gamma (B_d^{0, odd}),
\end{equation}
where $\Gamma (B_d^{0, even})$ ($\Gamma (B_d^{0, odd})$) is the width of $\Bd$
decay to the $CP$-even ($CP$-odd) final states, respectively. Assuming that
this difference is saturated by the $\Bd (\barBd) \to c \bar c d \bar d$ transition,
we obtain the following estimate:
\begin{eqnarray}
& & \sum_i (\textrm{Br}(B^0 \rightarrow f_i) S_i)  =  \nonumber \\
& & - \sin(2 \beta) [\textrm{Br}(B_d^{0, even}) - \textrm{Br}(B_d^{0, odd})] = \nonumber \\
& &  - \sin(2 \beta) \Delta \Gamma_d / \Gamma_d.
\label{dgamma}
\end{eqnarray}
%$\Delta \Gamma_d$ is due to decays of $B^0$ to
%flavor non-specific final states (to which both
%$B^0$ and $\bar{B}^0$ can decay).
In the SM framework 
%$\Delta \Gamma_d/\Delta m_d = \Delta \Gamma_s/\Delta m_s$. \cite{PDG}
%Substituting the measured values of $\Delta m_d$,
%$\Delta \Gamma_s$ and $\Delta m_s$ \cite{PDG} we obtain
%$\Delta	\Gamma_d = 0.01887 \pm 0.00026$.
$\Delta \Gamma_d / \Gamma_d = (42 \pm 8) \times 10^{-4}$ \cite{lenz}.
An alternative estimate of $\Delta \Gamma_d$ uses the SM
relation $\Delta \Gamma_d/\Delta m_d = \Delta \Gamma_s/\Delta m_s$
given in Ref. \cite{PDG}, page 1067. It results in a similar value 
$\Delta \Gamma_d / \Gamma_d = (44 \pm 6) \times 10^{-4}$.
Substituting the expression (\ref{dgamma}) in Eqs. (\ref{Ad_fi}) and (\ref{w3d-1}) we obtain
%$\sum_i{(\textrm{Br}(B^0 \rightarrow f_i) S_i)}$ by
%$- \sin{(2 \beta)} \Delta \Gamma_d / \Gamma_d$, and obtain
\begin{equation}
%A_S = (-0.064 \pm 0.022 \textrm{(tot)})\%.
A_S = (-0.045 \pm 0.016)\%.
\label{Ad_fi_3}
\end{equation}

{\bf Estimate 4.} In this estimate we use the measured value
$|\Delta \Gamma_d / \Gamma_d| = 0.015 \pm 0.018$ \cite{PDG}.
We replace $\sum_i (\textrm{Br}(B^0 \rightarrow f_i) S_i)$ by
$- \sin{(2 \beta)} \Delta \Gamma_d / \Gamma_d$
and obtain
\begin{equation}
A_S = (-0.16 \pm 0.20)\%,
\label{ASd_2}
\end{equation}

All these estimates of $A_S$ are consistent. In the following, we use the estimate
(\ref{Ad_fi_3}). Comparing it with the experimental result (\ref{Ac3}), we
conclude that CPV in interference of $\Bd$ decay with and
without mixing accounts for a part of the 
observed like-sign dimuon charge asymmetry. 
The experimental uncertainty on $\Delta \Gamma_d$ keeps open the possibility
of a substantially larger contribution from this source. 
We also note that
CPV in interference of $\Bd$ is much larger than the SM
prediction for CPV in mixing of $\Bd$ and $\Bs$ given in Eq. (\ref{SM_B0_mix}).

In Ref. \cite{D03} the like-sign
dimuon charge asymmetry is measured in several sub-samples of events
with an additional selection according to the muon impact parameter.
This selection effectively changes the contribution of muons coming
from the oscillated $\Bd$ decays. The estimate
(\ref{Ad_fi_3}) after applying this selection is also modified
and the contribution from CPV in interference can be enhanced by selecting
the dimuon events with large muon impact parameter.
Therefore, the study of the dependence of the asymmetry $A_S$
on the muon impact parameter can provide an additional insight on this
source of CPV. 

%These $A_S$ can be compared with (\ref{Ac3}).
%
%In conclusion, CPV in interference of decay with and
%without mixing of $B^0$ may account for the entire
%observed like-sign dimuon charge asymmetry.

\subsection{CPV in interference of $\Bs$ decay with and
without mixing}
% of $B_s$ to $f_j$, followed by $f_j \rightarrow \mu^- X$.}
%\textbf{4. CPV in interference of decay with and
%without mixing of $B_s$, followed by $\bar{c} \rightarrow \mu^- X$.}
Again we present several estimates.

{\bf Estimate 1.}
Four channels of interest are
%$D^+_s D^-_s$ with $\Gamma_i/\Gamma = (5.3 \pm 0.9) \times 10^{-3}$;
%$D^{*+}_s D^-_s + D^{*-}_s D^+_s$ with $\Gamma_i/\Gamma = (1.24 \pm 0.21) \%$;
$D^{(*)+}_s D^{(*)-}_s$ with $\Gamma_i/\Gamma = (4.5 \pm 1.4) \%$. \cite{PDG}
The CPV parameters $S_i$ have not been measured, but in the SM should
be approximately 
\begin{equation}
S_i = - \sin{(2 \beta_s)} \approx -0.036.
\end{equation}
Here, similarly to $\Bd \to D^{(*)+} D^{(*)-}$ decays, we assume that
%because these decays 
%contribute to $\Delta \Gamma_s$, and 
these final states are mainly CP-even. For these 4 decay channels we obtain
\begin{equation}
A_S(D^{(*)+}_s D^{(*)-}_s) = (-0.0003 \pm 0.0001 \textrm{(stat)})\%.
\label{ASd_s}
\end{equation}

{\bf Estimate 2.}
We assume that the final states $\bar{c} s c \bar{s}$
are mostly $CP$-even ($\eta_i = +1$), which is appropriate for 
$D^{(*)+}_{(s)} D^{(*)-}_{(s)}$ final states.
We take
\begin{equation}
\frac{\textrm{Br}(\Bs \rightarrow c \bar{c} s \bar{s})}
{(\textrm{Br}(b \rightarrow c \bar{c} s) +
 \textrm{Br}(b \rightarrow c \bar{c} d))}
\approx V_{cs}^2.
\end{equation}
%and $\eta_i = +1$, which is appropriate for $D^{(*)+}_{(s)} D^{(*)-}_{(s)}$.
Then Eq. (\ref{Ad_fi}) becomes approximately
\begin{eqnarray}
A_S & \approx & -0.5 w_3 f_s V_{cs}^2 \sin{(2 \beta_s)}
\frac{x_s}{1 + x_s^2}
%\frac{\Delta m_s \Gamma_s}{\Gamma_s^2 + \Delta m_s^2}
\frac{P_b - P_{\bar{b}}}{P_b P_{\bar{b}}} \nonumber \\
    & = & (-0.0013 \pm 0.0002 \textrm{(stat)})\%.
\label{As_fi_2}
\end{eqnarray}

The absolute value of estimate (\ref{As_fi_2}) 
can be considered as an upper bound on the contribution to $A_S$ 
from this source, because some of the $\Bs \to c \bar c s \bar s$ final states
are not $CP$-even.

{\bf Estimate 3.}
It is known that the four decay
channels $B_s \to D^{(*)+}_s D^{(*)-}_s$ do not exhaust
the contributions to $\Delta \Gamma_s$ \cite{PDG}.
To obtain an estimate of the like-sign dimuon charge asymmetry
from CPV in interference of $B_s$ we replace
$\sum_j \textrm{Br}(\Bs \rightarrow f_j) S_j$ by
$-\sin{(2 \beta_s)} \Delta \Gamma_s / \Gamma_s$.
We use the experimental value $\Delta \Gamma_s = 0.100 \pm 0.013$ ps$^{-1}$
and obtain
\begin{equation}
A_S(\Bs) = (-0.0009 \pm 0.0003 \textrm{(stat)})\%,
\label{ASs_2}
\end{equation}
which can be compared with (\ref{Ac3}).

In conclusion, CPV in interference of $\Bs$ decay with and
without mixing is suppressed by the
small values of $\sin{(2 \beta_s)}$ and $x_s / (1 + x_s^2)$. 

\subsection{Direct CPV in decay $\bar{b} \rightarrow c \bar{c} \bar{q}$ ($q = d$ or $s$),
followed by $\bar{c} \rightarrow \mu^- X$.}
%\textbf{5. CPV in decay $\bar{b} \rightarrow c \bar{c} \bar{q}$ ($q = d$ or $s$),
%followed by $\bar{c} \rightarrow \mu^- X$.}
This type of CPV occurs due to the interference of the tree level and
penguin diagrams with different strong phases and different weak phases.
Let us consider, as an example, the 
%form $(u \bar{b}) \rightarrow (u \bar{c}) (c \bar{d})$,
%namely 
decay $B^+ \rightarrow \bar{D}^0 D^+$, followed by
$\bar{D}^0 \rightarrow \mu^- X$. Its branching fraction is
$(3.8 \pm 0.4) \times 10^{-4}$. The CP-violating asymmetry in this decay has
not been measured, but should be less than $a = 0.1$ because
the penguin diagram is suppressed by one loop. The
like-sign dimuon charge asymmetry from this channel,
\begin{eqnarray}
A_S & = & a f_u \alpha \textrm{Br}(B^+ \rightarrow \bar{D}^0 D^+) \nonumber \\
& & \frac{\textrm{Br}(\bar{D}^0 \rightarrow \mu^- X)P_b 
- \textrm{Br}(D^+ \rightarrow \mu^+ X)P_{\bar{b}}}
{P_b P_{\bar{b}} \textrm{Br}(b \rightarrow \mu X)}, 
\end{eqnarray}
is less than 0.0002\%. Considering also
decays with $q = s$, which have a CP-violating asymmetry
suppressed by one loop $\times \lambda^2$, we conclude
that the direct $CP$ violation in the decays measured so far 
have a negligible contribution to the like-sign dimuon charge asymmetry.

\subsection{Direct CPV in semileptonic decays of $b$ and $c$ quarks.}
\label{semileptonic}
%\textbf{6. CPV in semileptonic decays of $b$ and $c$ quarks.}
In this subsection we assume that the like-sign dimuon charge
asymmetry is due solely to CPV in direct semileptonic decays
of charged and neutral hadrons containing $b$ or $c$ quarks.
This type of $CP$ violation vanishes in the lowest order due to the $CPT$ symmetry \cite{Gronau}. 
The second order calculations give extremely small value for the asymmetry $A_S$ of the
order of $10^{-9}$ \cite{Gronau}. Despite such a strong theoretical constraint, 
the possibility of large contribution from this source is discussed in Ref. \cite{Descotes}.
There is no direct experimental measurements
of this CPV in semileptonic $B$ decays, and the only experimental limitation 
can be derived from Eqs. (\ref{ac3}) and (\ref{Ac3}).
Let $a_{(b)}$ ($a_{(c)}$) be the flavor averaged
CP violating charge asymmetry of direct semileptonic
decays of $b$ ($c$) quarks.
The contribution of $a_{(b)}$ and $a_{(c)}$ to $a_S$ and $A_S$ are
\begin{eqnarray}
a_S & = & \frac{w_1 a_{(b)} + (w_2 + w_3 + w_5 + w_6) a_{(c)}}
  {w_1 + w_2 + w_3 + w_4 + w_5 + w_6}, \nonumber \\
    & = & 0.535 a_{(b)} + 0.454 a_{(c)},
\label{a_S_2}
\end{eqnarray}
and
\begin{eqnarray}
A_S & = & \frac{w_{1a} P_{\bar{b}} + w_{1b} P_b}{P_b P_{\bar{b}}} a_{(b)}+  \nonumber \\
    & & \frac{w_{2b} P_{\bar{b}} + w_{2a} P_b + 0.5 (w_3 + w_5) (P_{\bar{b}} + P_b)}
  {P_b P_{\bar{b}}} a_{(c)} \nonumber \\
    & = & 1.424 a_{(b)} + 0.525 a_{(c)}.
\label{A_S_2}
\end{eqnarray}
If the only asymmetry is $a_{(b)}$, then $A_S/a_S = 2.66$.
If the only asymmetry is $a_{(c)}$, then $A_S/a_S = 1.16$.
Taking $a_S$ from (\ref{ac3}), we obtain the following estimates for the contribution
of direct CPV to $A_S$:
\begin{eqnarray}
A_S (\mbox{from}~ a_{(b)}) & = & (-0.17 \pm 0.43) \%, \nonumber \\
A_S (\mbox{from}~ a_{(c)}) & = & (-0.07 \pm 0.19) \%.
\end{eqnarray}

Thus, the ``closure test" (\ref{ac3})
begins to constrain
the contributions of $a_{(c)}$
to the like-sign dimuon charge asymmetry $A_S$.

%% file: cpv_2012_conclusions.tex
\section{Conclusions}
\typeout{conclusions}
\begin{table}
\caption{\label{summary}
{Contributions to $A_S$ allowed by experiments.}
}
\begin{ruledtabular}
\newcolumntype{A}{D{A}{\pm}{-1}}
\newcolumntype{B}{D{B}{-}{-1}}
\begin{tabular}{lll}
   & Process & allowed $A_S$ \\
\hline
%A & Mixing of $B^0$ and $B_s$ & $(+0.000 \pm 0.012 \textrm{ (stat)})\%$ \\
A & Mixing of $\Bd$ & $(+0.062 \pm 0.073 )\%$ \\
A & Mixing of $\Bs$ & $(-0.111 \pm 0.093 )\%$ \\
%B & Mixing of $B^0$ & $(0.11 \pm 0.11 \textrm{ (stat)})\%$ \\
%B & Interference of $B^0$ & $(-0.064 \pm 0.022 \textrm{ (tot)})\%$ \\
B & Interference of $\Bd$ & $(-0.045 \pm 0.016 )\%$ \\
C & Interference of $\Bs$ & $(-0.0009 \pm 0.0003)\%$ \\
D & CPV in $b \to c \bar{c} \bar{q}$ decays & $(+0.000 \pm 0.001)\%$ \\
E & $a_{(b)}$  in $b \to \mu X$ decays & $(-0.17 \pm 0.43 )\%$ \\
E & $a_{(c)}$  in $c \to \mu X$ decays & $(-0.07 \pm 0.19 )\%$ \\
%F & Production of $c \bar{b}$ & $(-0.01 \pm 0.03 \textrm{ (stat)})\%$ \\
\end{tabular}
\end{ruledtabular}
\end{table}

We have considered several possible causes of
the measured like-sign dimuon charge asymmetry $A_S$,
and obtained their experimental constraints.
A summary is presented in Table \ref{summary}.
We find that standard model CP violation in interference of decays with and
without mixing of $B^0$ to flavor
non-specific states $f_i$,
followed by the decay $f_i \rightarrow \mu X$,
contributes
\begin{equation}
%A_S(B^0) = (-0.064 \pm 0.022 \textrm{ (tot)})\%
A_S(B^0) = (-0.045 \pm 0.016 \textrm{(stat)})\%
\label{As_B0}
\end{equation}
to the like-sign dimuon charge asymmetry $A_S$.
CP violation in interference does not contribute to
the inclusive muon charge asymmetry and therefore
is compatible with the observation that $a_S$
is consistent with zero. 

Among all other possible sources of 
the dimuon charge asymmetry, only the direct $CP$ violation
in semileptonic $b$- and $c$-hadron decays is not yet limited experimentally.
It is very small in the SM, but, until experimentally measured, this source
of dimuon charge asymmetry cannot be excluded.
Our estimate of this source is derived from the D\O\ measurements (\ref{ac3}) and (\ref{Ac3}).
The exclusive measurement of this type of $CP$ violation is required
to improve this constraint.

Taking into account the additional SM source of dimuon charge asymmetry 
(\ref{As_B0}) identified 
in this paper, the combination of D\O\ measurements (\ref{ac3}) and (\ref{Ac3})
becomes consistent with the SM expectation within 3 standard deviations.
Still the difference between (\ref{Ac3}), and (\ref{SM_B0_mix}) and (\ref{As_B0}),
$(-0.32 \pm 0.14 \textrm{ (tot)})\%$,
leaves some room for new physics $CP$ violation in {\Bd} and {\Bs} mixing,
in the interference of {\Bd} and {\Bs} decays with and without mixing, or
in semileptonic decays of $b$ and $c$ hadrons. 
A deviation in the value of $\Delta \Gamma_d$
from the SM prediction could also contribute to the difference between the
observed and expected like-sign dimuon charge asymmetry.
%production of $c \bar{b}$,
%or a statistical fluctuation.

%% file: cpv_2012_acknowledgement.tex
% acknowledgement_paragraph_r2.tex                24 March 2010
%
As D0 collaborators deeply involved in the dimuon asymmetry analysis,
we thank the D\O\ Collaboration for inspiring, commenting, and supporting this work.
We thank Michael Gronau and Jonathan L. Rosner for very useful comments 
and suggestions to our paper. 

%% file: cpv_2012_main_PRD.bbl
\begin{thebibliography}{99}
\bibitem{D01}
V.M.~Abazov \textit{et al.} (D0 Collaboration), Phys. Rev. {\bf D 74}, 092001 (2006).

\bibitem{D02}
V.M.~Abazov \textit{et al.} (D0 Collaboration), Phys. Rev. {\bf D 82}, 032001, (2010).

\bibitem{D02a}
V.M.~Abazov \textit{et al.} (D0 Collaboration), Phys. Rev. Lett. {\bf 105}, 081801 (2010).

\bibitem{D03}
V.M.~Abazov \textit{et al.} (D0 Collaboration), Phys. Rev. {\bf D 84}, 052007 (2011).

\bibitem{Grossman}
Y.~Grossman {\it et al.}, Phys.~Rev.~Lett.~{\bf 97}, 151801 (2006).

\bibitem{lenz}
A. Lenz and U. Nierste, : in Proceedings of CKM2010, Warwick, 2010, edited by Tim
Gershon, eConf C100906 (2010), arXiv:1102.4274 [hep-ph].

\bibitem{hfag}
Y. Amhis et al. (Heavy Flavor Averaging Group), arXiv:1207.1158 [hep-ex] and
online update at http://www.slac.stanford.edu/xorg/hfag

%\bibitem{Descotes} S\'{e}bastien Descotes-Genon and Jernej F. Kamenik,
%arXiv:1207.4483 (2012).

%\bibitem{pythia} T.~Sj\"{o}strand {\it et al.}, Comput.~Phys.~Commun.~{\bf 135}, 238 (2001).

\bibitem{PDG}
Review of Particle Physics, J. Beringer et al. (Particle Data Group),
Phys. Rev. {\bf D 86}, 010001 (2012).

%\bibitem{BH}
%B. Hoeneisen, Proceedings of the DPF-2011 Conference,
%arXiv:1109.1438 [hep-ex] (2011).

%\bibitem{CPV}
%Gustavo Castelo Branco, Lu\'{\i}s Lavoura and Jo\~{a}o Paulo Silva,
%\textit{CP Violation}, Oxford Science Publications (1999).

%\bibitem{Randall}
%L. Randall and S. Su, Nucl. Phys. {\bf B540}, 37 (1999).

%\bibitem{Hewett}
%J. L. Hewett, hep-ph/9803370.


%\bibitem{pdg}
%K. Nakamura \textit{et al.} (Particle Data Group),
%Journal of Physics G~\textbf{37} 075021 (2010).

%\bibitem{sakharov}
%A.D.~Sakharov, Pisma~Zh.~Eksp.~Teor.~Fiz.~{\bf 5}, 32 (1967)
%[Sov.~Phys.~JETP~Lett.~{\bf 5}, 24 (1967)].

%\bibitem{newcp}
%M.B.~Gavela, {\it et al.}, Mod. Phys. Lett. A{\bf 9}, 795 (1994); \\
%M.B.~Gavela, {\it et al.}, Nucl. Phys. B~{\bf 430}, 382 (1994); \\
%P.~Huet and E.~Sather, Phys. Rev. D~{\bf 51}, 379 (1995).

%\bibitem{babar}
%B.~Aubert \textit{et al.} (Babar Collaboration), Phys. Rev. Lett. \textbf{96}, 251802 (2006); \\
%B.~Aubert \textit{ et al.} (Babar Collaboration), arXiv:hep-ex/0607091 (2006).

%\bibitem{belle}
%E.~Nakano \textit{et al.} (Belle Collaboration), Phys.~Rev.~D~\textbf{73}, 112002 (2006).

%\bibitem{Randall}
%L.~Randall and S.~Su, Nucl.~Phys.~B~\textbf{540}, 37 (1999).

%\bibitem{Hewett}
%J.L.~Hewett, arXiv:hep-ph/9803370 (1998).

%\bibitem{Hou}
%G.W.S.~Hou, arXiv:0810.3396 [hep-ph] (2008).

%\bibitem{Soni}
%A.~Soni {\it et al.}, Phys.~Lett.~B~\textbf{683}, 302 (2010); \\
%A.~Soni {\it et al.}, arXiv:1002.0595 (2010) [hep-ph] and references therein.

%\bibitem{buras}
%M.~Blanke, {\it et al.}, JHEP {\bf 0612}, 003 (2006); \\
%W. Altmannshofer, {\it et al.}, Nucl. Phys. B {\bf 830}, 17 (2010).

%\bibitem{charge}
%Charge conjugation invariance is implied throughout this article.

%\bibitem{run2muon} V.M.~Abazov {\it et al.},
%Nucl.~Instrum.~Methods in Phys.~Res.~A~{\bf 552}, 372 (2005).

%\bibitem{run2det}
%V.M.~Abazov {\it et al.} (D0 Collaboration), Nucl.~Instrum.~Methods  in Phys.~Res.~A~{\bf 565}, 463 (2006).

%\bibitem{layer0}
%S.N.~Ahmed {\it et al.}, arXiv:1005.0801 [physics.ins-det] (2010), accepted for
%publication in Nucl.~Instrum.~Methods  in Phys.~Res.~A.; \\
%R.~Angstadt {\it et al.}, arXiv:0911.2522 [physics.ins-det] (2009), submitted for
%publication in Nucl.~Instrum.~Methods  in Phys.~Res.~A.

%\bibitem{rapidity}
%Pseudorapidity $\eta \equiv -\ln{ \left[ \tan(\theta/2) \right] }$, where
%$\theta$ is the polar angle of the track relative to the direction of the
%proton beam; $\phi$ is the azimuthal angle with respect to the proton beam,
%and $\phi = 90^0$ is defined as the vertical axis.

%\bibitem{evtgen} D.G.~Lange, Nucl. Instrum. Methods in Phys. Res. A~{\bf 462}, 152 (2001);
%for details see {\sc http://www.slac.stanford.edu/\verb"~"lange/EvtGen}.

%\bibitem{pdf} J.~Pumplin {\it et al.}, J.~High~Energy~Phys.~{\bf 0602}, 032 (2006).

%\bibitem{geant}
%R.~Brun and F.~Carminati, CERN program library long writeup W5013 (unpublished).

%\bibitem{mu}
%In this section the symbol ``$\mu$" and word ``muon" stand for
%``muon candidate passing a specified set of selections."

%\bibitem{bjk}
%V.M.~Abazov \textit{et al.} (D0 Collaboration), \\ Phys.~Rev.~Lett.~\textbf{100}, 211802 (2008).

%\bibitem{hfag}
%E.~Barberio {\it et al.} (HFAG), arXiv:0808.1297 [hep-ex] (2008).

%\bibitem{asl-d0}
%V.M.~Abazov \textit{et al.} (D0 Collaboration), arXiv:0904.3907 [hep-ex], accepted for
%publication in Phys.~Rev.~D.

%\bibitem{D0-phi}
%V.M.~Abazov \textit{et al.} (D0 Collaboration), Phys. Rev. Lett.~\textbf{101}, 241801 (2008); \\
%V.M.~Abazov \textit{et al.} (D0 Collaboration), Phys. Rev. Lett.~\textbf{98}, 121801 (2007).

%\bibitem{CDF-phi}
%%T.~Aaltonen \textit{et al.} (CDF Collaboration), CDF Public Note CDF/ANAL/BOTTOM/PUBLIC/9458 (2008); \\
%T.~Aaltonen \textit{et al.} (CDF Collaboration), Phys. Rev. Lett.~\textbf{100}, 161802 (2008).

%\bibitem{5928}
%CDF/D\O\ $\Delta \Gamma_s$, $\beta_s$
%Combination Working Group, D\O\ Note 5928-CONF (2009), and
%CDF Public Note CDF/ANAL/BOTTOM/PUBLIC/9787 (2009).

%\bibitem{CPV}
%G.C.~Branco, L.~Lavoura and J.P.~Silva, {\it ``CP Violation"} (Clarendon, Oxford, 1999).


\bibitem{asld-d0}
V.M.~Abazov \textit{et al.} (D0 Collaboration), Phys. Rev. {\bf D 86}, 072009 (2012).

\bibitem{asls-d0}
V.M.~Abazov \textit{et al.} (D0 Collaboration), Phys. Rev. Lett. {\bf 110}, 081801 (2010).

\bibitem{asls-lhcb}
LHCb collaboration, Conference report LHCb-CONF-2012-022 (2012).

\bibitem{Nierste}
A.~Lenz and U.~Nierste, J.~High~Energy~Phys.~{\bf 0706}, 072 (2007).

%\bibitem{CKMfitter} S. Descotes-Genon (on behalf of the CKMfitter group),
%arXiv:1209.4016 (2012).

\bibitem{lenz2}
A. Lenz, arXiv:1106.3200 (2012).

\bibitem{Sanda}
A. I. Sanda and Z. Z. Xing, Phys. Rev. {\bf D 56}, 341 (1997).

\bibitem{Xing1}
Z. Z. Xing, Phys. Lett. {\bf B443}, 365 (1998).

\bibitem{Xing2}
Z. Z. Xing, Phys. Rev. {\bf D 61}, 014010 (1999).

%\cite{Dunietz:2000cr}
\bibitem{Dunietz}
  I.~Dunietz, R.~Fleischer and U.~Nierste,
  %``In pursuit of new physics with $B_s$ decays,''
  Phys.\ Rev.\ {\bf D 63} (2001) 114015 [arXiv:hep-ph/0012219].
  %%CITATION = HEP-PH/0012219;%%

%\bibitem{lenz1}
%A. Lenz, arXiv:1205.1444 [hep-ph].

%\bibitem{LHCb}
%LHCb collaboration, A. Bharucha et.al., arXiv:1208.3355 (2012)

%\bibitem{WA} Wine and Cheeze by Mark Williams, 19th October 2012. {\color{red}(Will be updated.)}
\bibitem{Gronau}
S.~ Bar-Shalom, G.~ Eilam, and M. Gronau, 
Phys. Lett. B {\bf 694}, 374 (2011) and arXiv:1008.4354 [hep-ph].

\bibitem{Descotes}
S. Descotes-Genon and J.F.~ Kamenik, arXiv:1207.4483 [hep-ph].


\end{thebibliography}
